\documentclass[a4paper,12pt]{article}
\providecommand{\keywords}[1]{\textbf{Keywords:} #1}
\usepackage{amsmath}
\usepackage{subcaption}
\usepackage{amssymb}
\usepackage{amsthm}
\usepackage{authblk}
\usepackage{geometry}
\usepackage{graphicx}
\usepackage{hyperref}

\newtheorem{example}{Example}
\title{ A simulation study comparing statistical approaches for estimating extreme quantile regression with an application to forecasting of fire risk }
\author{Amina EL BERNOUSSI and Mohamed EL ARROUCHI }
\affil{\emph{ Department of Mathematics, Faculty of Sciences, Ibn Tofail University, Kenitra, Morocco}\\
 \emph{E-mail : amina.elbernoussi@uit.ac.ma}}
\date{}

\begin{document}

\maketitle

\begin{abstract}
 This simulation study compares statistical approaches for estimating extreme quantile regression, with a specific application to fire risk forecasting. A simulation-based framework is designed to evaluate the effectiveness of different methods in capturing extreme dependence structures and accurately predicting extreme quantiles. These approaches are applied to fire occurrence data from the Fez-Meknes region, where a positive relationship is observed between increasing maximum temperatures and fire frequency. The study highlights the comparative performance of each technique and advocates for a hybrid strategy that combines their complementary strengths to enhance both the accuracy and interpretability of forecasts for extreme events.
\end{abstract}
\section{Introduction}
Recently, quantile regression has drawn a lot of interest from both an empirical and theoretical perspective. Quantile regression, to put it simply, is a statistical technique used to estimate conditional quantile functions. Quantile regression techniques are based on minimizing asymmetrically weighted absolute residuals and are designed to estimate conditional median functions as well as a wide range of other conditional quantile functions. This is comparable to classical linear regression techniques, which are based on minimizing sums of squared residuals and intended to estimate models for conditional mean functions. The main reason for utilizing quantiles instead of simple mean regression is that it is more easier and more accurate to depict the stochastic relationship between random variables. According to a thorough discussion in Koenker and Bassett \cite{ko}, quantile regression would yield more reliable and, as a result, more effective estimates in linear models built on certain non-Gaussian settings, where traditional least squares estimators may be gravely inadequate.
Furthermore, the benefit at the non-Gaussian spreads would not be greater than the cost in terms of efficiency loss (to the least squares estimators) at the normal distribution. Therefore, quantile regression techniques enhance and supplement conventional means regression models.\\

 A regression-type model is presented in the Carvalho's et al. \cite{ca} study for scenarios in which the covariates and the response are multivariate extreme values. In contrast to conventional regression techniques, this model considers the fact that compound maxima's limiting distribution is an extreme value copula. The regression variety, which is a family of regression lines that obey this asymptotic result, is a key aim in this approach. The authors employ a prior Bernstein polynomial on the space of angular densities to learn the suggested model from data, resulting in an induced prior on the space of regression varieties. The suggested methods perform well in numerical simulations, and an example using actual financial data highlights intriguing features of the conditional risk of large losses in two significant global stock markets. In this context, El Bernoussi and El Arrouchi \cite{el} introduces a novel regression model specifically tailored to handle non-stationary extremes by relying on block maxima for both the response and covariate variables, incorporating extreme quantile regression in a bivariate case. To better capture the dependence structure between these extremes, the authors propose a prior distribution on spectral densities using a Logistic-Normal distribution, which offers greater flexibility and modeling capacity than the traditionally employed Dirichlet distribution.\\

Examples of classical approaches include kernel methods (Daouia et al. \cite{da}; Gardes and Stupfler \cite{ga}; Velthoen et al. \cite{ve19}), generalized additive models (Chavez-Demoulin and Davison \cite{cha}; Youngman \cite{you}), and linear models (Wang et al. \cite{wa}; Li and Wang \cite{li}). However, according to Gnecco et al. \cite{gn}, those methods usually have drawbacks in terms of covariate dimensionality or flexibility because they are not made to capture non-linear or non-additive interactions. This sparked a recent interest in more adaptable and machine-learning based methods to deal with more complicated multivariate dependencies, such as neural networks (Pasche and Engelke \cite{pa}; Richards and Huser \cite{ri}; Allouche et al. \cite{al}) or tree-based ensembles (Velthoen et al. \cite{ve23}; Gnecco et al \cite{gn}; Koh \cite{koh}). Even for static estimates, it might be challenging to provide trustworthy confidence intervals (CIs) for extreme quantiles (Zeder et al. \cite{ze}). In order to estimate covariate dependent CIs for complicated dependence models, bootstrap approaches are typically used (Davison and Hinkley \cite{dav97}; Davison et al. \cite{dav03}). For some extreme statistics, bootstrap CIs are well understood (de Haan and Zhou \cite{deh}), but not for the more adaptable covariate-dependent models. The sensitivity of extreme value estimations to the most extreme observations is one of the primary challenges. For instance, non-parametric bootstrap statistics may display undesirable discontinuous behaviours on finite samples as a result of this.\\
 The Extreme Quantile Regression Neural Network (EQRN) model, which combines neural networks and extreme value theory to estimate high quantiles beyond historical evidence, is presented by Pasche et al. \cite{pa}. By employing spatiotemporal covariates to predict return levels and exceedance probabilities at one-day intervals, the EQRN model—specifically, its recurrent version—predicts the danger of floods in Switzerland and adjusts forecasts to climatic fluctuations. The intricacy of temporal connections has been successfully captured by this model, which may find use in other domains where conditional extreme quantile forecasts are necessary.\\

Pasche et al. \cite{pa} compared their extreme quantile regression neural (EQRN) algorithm with the previously described techniques in their study. For this reason, we decide to contrast Carvalho's et al. \cite{ca} approach with Pasche et al. \cite{pa}.
A numerical simulation demonstrates the superiority of the EQRN model over the Carvalho et al.'\cite{ca} model in terms of quantile prediction level efficiency. This demonstrates how accurately the risk has been predicted.\\
The paper's sections are as follows: Section \ref{sec2} covers the theoretical underpinnings of the two approaches and their numerical simulation. Then, the simulation study and it results examinates the outcomes to properly assess how well the two techniques performed. We wrap up with an application to estimate the fire risk in Morocco's Fez-Meknes region by taking into account the maximum temperatures from five stations: Ifrane, Boulemane, Fez, Sefrou, and Meknes-EL Hajeb, as covariates, and the total number of fires in the area as a response variable. We conclude with a discussion in section \ref{sec3}.

\section{Methodology}\label{sec2}
 In this section, we describe our work, which will be presented as follows. The theoretical underpinnings of the two quantile regression models that the two authors Carvalho et al. \cite{ca} and Pasche et al. \cite{pa} introduced are first presented. The first involves estimating the spectral measure and applying the Generalized Extreme Value (GEV) method. The second is the neural network-based Multivariate Peaks-Over-Threshold (POT) Method.
 In order to compare each's performance in quantile estimation and use Root-Mean-Square Error (RMSE), and Mean Absolute Error (MAE) to estimate risk, we then access a numerical simulation. Next, we select the most effective model for calculating the fire risk in Morocco's Fez-Meknes region. 
\subsection{Background on the quantile regression}
It is demonstrated that a straightforward minimization problem that produces the location model's ordinary sample quantiles automatically generalizes to the linear model, producing a new class of statistics that we refer to as "regression quantiles." One significant special example is the estimator that minimizes the sum of the absolute residuals. The combined asymptotic distribution of regression quantiles and a few equivariance features are determined. These findings allow for a straightforward extension of some well-known robust estimators of location to the linear model. 
Estimators that significantly outperform the least-squares estimator over a broad class of non-Gaussian error distributions are proposed, with equivalent efficiency to least squares for Gaussian linear models. 
The initial version of quantile regression, proposed by Koenker and Bassett in 1978, involves linearly modeling the conditional quantile of a response $Y$ given a covariate $X=(X_1,…,X_p)^T$ that is, 
\begin{equation}\label{eq0}
  F^{-1}(q|x)=x^T\beta_q, \,\,\,\,\,\,\,\, 0<q<1,
\end{equation}
where $F^{-1}(q|x)=\inf\{y:F^{-1}(q|x)\geq q\}$ and $F(y|x)$ is the distribution function of $Y|X=x$.\\

 Carvalho et al. \cite{ca} work with a regression of block maxima on block maxima and they focus on the bivariate case. Let $\{(X_i,Y_i)\}_{i=1}^n$ be a sequence of independent random vectors with unit Fréchet marginal distributions, i.e $\exp(-1/z)$ for $z>0$. $Y_i$ is considered as a response but $X_i$ is a p-dimensional covariate. Let the componentwise block maxima defined as follows : $ M_n=(M_{n,x_1},...,M_{n,x_p},M_{n,y})$ with $M_{n,y}=\max(Y_1,...,Y_n)$ and $M_{n,x_j}=\max(X_{j,1},...X_{j,p})$ for $j=1,...,p.$
 It is commonly known that in this configuration, the vector of normalized componentwise maxima $M_n/n$ converges in distribution to a random vector $( X,Y)$ which follows a multivariate extreme value distribution with the joint distribution function
 \begin{equation}\label{eq1}
   G(x,y)=\exp\{-V(x,y)\}, \,\, \, x\in (0,\infty)^{p+1}
 \end{equation}
with \begin{equation}\label{eq2}
       V(x,y)=d \int_{\Delta_d}\max\left(\frac{w_1}{x_1},...,\frac{w_p}{x_p},\frac{w_{p+1}}{y}\right)H(dw)
     \end{equation}
is the exponent measure and $d=p+1.$ Furthermore, the angular measure, or $H$, is a parameter of the multivariate extreme value distribution G that regulates the dependence between the extreme values; in particular, $H$ is a probability measure on the unit simplex $\Delta_d=\{(w_1,...,w_d)\in [0,1]^d,\sum_{i=1}^d w_i=1\}\subset \mathbb{R}^d,$ and 
obeying the mean constraint 
\begin{equation}\label{eq3}
  \int_{\Delta_d}wH(dw)=\frac{1}{d}\textbf{1}_d,
\end{equation} 
where $\textbf{1}_d=(1,...,1)\in \mathbb{R}^d$. In the event where H is absolutely continuous with respect to the Lebesgue measure, the Radon–Nikodym derivative provides its density $h=dH/dw,$ for $w\in \Delta_d$.
They define the family of regression lines 
\begin{equation}\label{eq4}
  \emph{L}=\{\textbf{L}_q: 0<q<1\} , \,\,\,\,\,\,\, \textbf{L}_q=\{y_{q|x} : x\in (0,\infty)^p\}
\end{equation}
where 
\begin{equation}\label{eq5}
 y_{q|x}=\inf\{y>0: G_{Y|X}(y|x)\geq q\}
\end{equation}
is a conditional quantile of a multivariate extreme value distribution, with $q\in (0, 1)$ and $x\in (0,\infty)^p$ , and $G_{Y|X}(y|x)=\mathbb{P}(Y \leq y |X=x)$ is a conditional multivariate 
extreme value distribution function. In bivariate case, $G_{Y|X}(y|x)=\mathbb{P}(Y \leq y |X=x)$ is a conditional bivariate extreme value distribution function as
 defined by 
 \begin{equation}\label{eqaz}
   G_{Y|X}(y|x)=2\exp\left\{-2\int_{0}^{1} \max\left(\frac{w}{x},\frac{1-w}{y}\right)h(w)dw+x^{-1}\right\}\int_{w(x,y)}^{1}wh(w)dw.
 \end{equation}
with $w(x,y)=x/(x+y)$ and $x,y>0$.\\
 We will now examine certain parametric instances of regression manifolds as recently defined.
\begin{example}(Logistic model)
This follows from the Logistic bivariate extreme value distribution function given by
\begin{equation*}
  G(x,y)=\exp\{-(x^{-1/\alpha}+y^{-1/\alpha})^\alpha\}, x,y > 0
\end{equation*}
where $\alpha \in (0,1]$ represents the strength of dependence between extremes. As $\alpha$ approaches 0,
 the dependence becomes stronger, with $\alpha \rightarrow 0$ indicating perfect dependence. The conditional
 distribution of $Y$ given $X$ is
\begin{equation*}
  G_{Y|X}(y|x)= G(x,y)(x^{-1/\alpha}+y^{-1/\alpha})^{\alpha-1}x^{1-1/\alpha}\exp(1/x),x,y> 0,
\end{equation*}
For regression manifolds representation, we use the approximation of the exact logistic regression manifolds defined by Carvalho et al. \cite{ca} as follows for $q \in (0,1)$ and $x>> 1$
\begin{equation}\label{eqlog}
  \tilde{y}_{q|x}=\frac{\alpha}{1-\alpha}\{q^{1/(\alpha-1)}-1\}^{-\alpha-1}\{q^{\alpha/(1-\alpha)}- 1\}q^{1/(\alpha-1)} + \{q^{-1/(1-\alpha)}- 1\}^{-\alpha}x
\end{equation}
\end{example}
\begin{example}( Husler-Reiss model)
  This follows from the Husler-Reiss bivariate extreme value distribution function given by
  \begin{equation*}
    G(x,y) = \exp \left\{-x^{-1}\Phi\left(\lambda+\frac{1}{2\lambda}\log\frac{y}{x}\right)-y^{-1}\Phi\left(\lambda+\frac{1}{2\lambda}\log\frac{x}{y}\right) \right\}, x, y > 0
  \end{equation*}
where $\Phi$ denotes the standard normal distribution function and $\lambda \in (0,\infty]$ is a parameter that
 modulates the dependency between extremes: as $\lambda\rightarrow 0$, the dependence becomes perfect, while in
 the limit as $\lambda\rightarrow \infty$, the dependency reaches complete independence. The collection of regression
 lines $\mathbf{L}_q$ for this model lacks explicit forms and is derived using \ref{eq5} with
 \begin{equation*}
   G_{Y|X}(y|x)=\left[\Phi \left(\lambda+\frac{1}{2\lambda}\log\frac{y}{x}\right)+\frac{1}{2\lambda}\phi\left(\lambda+\frac{1}{2\lambda}\log\frac{y}{x}\right)-\frac{xy^{-1}}{2\lambda}\phi\left(\lambda+\frac{1}{2\lambda}\log\frac{x}{y}\right)\right]\times G(x,y)\exp(1/x)
 \end{equation*}
  for $x,y > 0$ where $\phi$ is the standard Normal density function.
\end{example}
\begin{example}(Coles-Tawn model)
 This follows from the Coles-Tawn bivariate extreme value distribution function given by
 \begin{equation*}
   G(x,y)=\exp[-x^{-1}\{1-Be(q;\alpha+1,\beta)\}-y^{-1}Be(q;\alpha,\beta+1)],x,y>0
 \end{equation*}
 where $Be(q;a,b)$ represents the cumulative distribution function of a Beta distribution with parameters $a,b > 0$. In this context, $q$ is defined as $q=\alpha y^{-1}/(\alpha y^{-1}+\beta x^{-1})$ with $\alpha,\beta > 0$ serving as
 parameters that control the dependence between extremes. The scenario where $\alpha = \beta = 0$ indicates
 complete independence, while the case of $\alpha=\beta \rightarrow \infty$ signifies perfect dependence. For fixed values
 of $\alpha(\beta)$, an increase in $\beta(\alpha)$ leads to a stronger dependence. The family of regression lines $\mathbf{L}_q$ for
 this model does not have a straight forward representation and is determined through calculations
 using \ref{eq5} for $x,y > 0$, with
 \begin{equation*}
   G_{Y|X}(y|x)=\left[1-Be(q;\alpha+1,\beta)+\frac{(\alpha+1)\beta}{\gamma}be(q;\alpha+2,\beta+1)-\frac{x}{y}\frac{\alpha(\beta+1)}{\gamma}be(q;\alpha+1,\beta+2)\right]
 \end{equation*}
 \begin{equation*}
   G(x,y)\exp(1/x).
 \end{equation*}
\end{example}

As per Hanson et al. \cite{ha}, they use a Bernstein polynomial defined on the unit simplex $\Delta_d$ to characterize the angular density $h$; as a result, basis polynomials are Dirichlet densities. More specifically, their angular density specification is 
\begin{equation}\label{eq6}
  h(w)=\sum_{|\alpha|=J} \pi_\alpha dir_d(w;\alpha)
\end{equation}
with $w\in \Delta_d$ and $\pi_J=(\pi_\alpha : |\alpha|=J).$ The dirichlet density $dir_d$ is defined as 
$$dir_d(w;\alpha)=\frac{\Gamma(|\alpha|)}{\prod_{i=1}^d \Gamma(\alpha_i)}\prod_{i=1}^d w_i^{\alpha_i -1}$$
where $\alpha \in \mathbb{N}^d$, $|\alpha|=\sum_{j=1}^{d} \alpha_j$, $\Gamma(z)=\int_{0}^{\infty} x^{z-1} \exp(-x)dx$ is the gamma function, $\pi_\alpha>0$ are weights and $J\in \mathbb{N}$ controls the order of the resulting polynomial. The weights must obey 
\begin{equation}\label{eq7}
  \sum_{|\alpha|=J} \pi_\alpha =1, \,\,\,\,\,\,\,\, \sum_{i=1}^{J-d+1} i \sum_{|\alpha|=J,\alpha_j=i} \pi_\alpha=\frac{J}{d}, \,\,\,\, j=1,...,d 
\end{equation}
There are $m-d$ parameters, according to the normalization and mean constraints in \ref{eq7}, where $m=\binom {J-1} {d-1}$ is the number of basis functions in \ref{eq6}; these free weights are represented as $\{\pi_\alpha: \alpha \in \texttt{F}\}$, where $\texttt{F}=\{ \alpha \in \mathbb{N}^d: |\alpha|=J, \alpha \notin\{ a_1,…, a_d\}\}$, with $j$ being a $J$-vector of ones, with the exception that element $i$ is $J-d+1$. Like Hanson et al. \cite{ha}, they use a generalized logit transformation to parametrize the free weights, which implicitly specifies the auxiliary parameters $\pi'_\alpha$,
\begin{equation}\label{eq8}
  \pi_\alpha=\frac{\exp(\pi'_\alpha)}{d+\sum_{\tilde{\alpha}\in \texttt{F}}\exp(\pi'_{\tilde{\alpha}})}
\end{equation}
The angular density in \ref{eq6} results in an induced prior on the space of regression lines $\textbf{L}_q$ after integration with respect to y and inversion of $G_{Y|X} (y|x)$. This is done in order to induce a prior in the space of regression manifolds. In particular, they follow these steps to define a prior on the space of regression manifolds. A prior on the space of regression lines $\textbf{L}_q=\{y_{q|x}: x\in (0,\infty)\}$ is induced by the Bernstein polynomial prior in \ref{eq6}, where $y_{q|x}$ is a solution to the equation $G_{Y|X} (y|x)=q$, for $q\in (0,1)$, where 
\begin{equation*}
  G_{Y|X} (y|x)
  =\frac{2}{J}\exp\left\{-\frac{2}{J} \sum_{|\alpha|=J}\pi_\alpha[\alpha_1x^{-1}\{1-Be(w(x,y);\alpha_1+1,\alpha_2)\}+\alpha_2y^{-1}Be(w(x,y);\alpha_1,\alpha_2+1)]\right\}
\end{equation*}
    
\begin{equation}\label{eq9}
\times \sum_{|\alpha|=J}\pi_\alpha \alpha_1 \{1-Be(w(x,y);\alpha_1+1,\alpha_2)\}\exp(1/x),
\end{equation}
 where $w(x,y)=x/(x+y)$ for $x,y>0$. Lastly, they establish the following Dirichlet prior on the free parameters
  \begin{equation*}
    p(\pi_\alpha:\alpha\in \texttt{F})\approx dir_d(w|c\textbf{1}_m)\prod_{j=1}^{d}I\left\{\sum_{i=1}^{J-d+1} i \sum_{|\alpha|=J,\alpha_j=i} \pi_\alpha =\frac{J}{d}\right\}
  \end{equation*}
  
  where $I$ is the indicator function, to finish the model definition. This creates a prior on the auxiliary parameters $\pi'_\alpha$ in \ref{eq8}, for more details see Carvalho et al. \cite{ca}.\\
  
  Pasche et al. \cite{pa} propose the extreme quantile regression neural networks (EQRN) model, an approach that can extrapolate in the presence of complicated predictor dependence by fusing techniques from extreme value theory and neural networks. They use a conditional version of the generalized Pareto distribution (GPD)
  \begin{equation}\label{eq10}
   \mathbb{P}(Y>y|X=x)\approx (1-\tau_0)\left(1+\xi(x)\frac{y-u(x)}{\sigma(x)}\right)_{+}^{-1/\xi(x)}\,\,\,\,\,\,\, ,\,\,\,y>u(x)
  \end{equation}
  where the shape $\xi(x) \in \mathbb{R}$ and scale $\sigma(x)>0$ rely on the covariates, and the threshold $u(x)$ is selected as an intermediate quantile $Q_x(\tau_0)$ at level $\tau_0\in(0,1)$ near $1$; in this case, we ignore the dependency of $\sigma(x)$ on the intermediate level $\tau_0$ in the notation. For the exact assertion, see Pickands III \cite{pi} and Balkema and de Haan \cite{ba}; this approximation is valid under weak conditions on the tail of $Y|X = x$. Since this condition is univariate, it can be checked even in more complicated cases, such as when $X$ is the history of a multivariate time series.
 Since it encodes the response's tail heaviness, the shape parameter $\xi(x)$ is crucial. If it is positive, the response has a heavy-tailed distribution like Pareto or Student-t; if it is zero, it has a light-tailed distribution like Gaussian or exponential; and if it is negative, it has a finite upper endpoint. Using approximation \ref{eq10}, we can invert this expression to forecast an extreme quantile at level $\tau>\tau_0$
 \begin{equation}\label{eq11}
   Q_x(\tau):= Q_x(\tau_0)+\frac{\sigma(x)}{\xi(x)}\left[\left(\frac{1-\tau_0}{1-\tau}\right)^{\xi(x)}-1\right]
 \end{equation}
 This demonstrates that estimates of the intermediate quantile, $\tilde{Q}_x(\tau_0)$, and the conditional GPD parameters, $\tilde{\xi}(x)$ and $\tilde{\sigma}(x)$, as functions of the predictor vector, are necessary for an estimate $Q_x(\tau)$ of an extreme quantile. Any of the already available quantile regression techniques can be applied to the intermediate quantile function since, as was previously said, they are effective for this moderate quantile level. One can specify a parametric or nonparametric model to estimate the GPD parameters. The EQRN approach provides precise estimates for quantile functions $Q_x(\tau)$ at extreme levels $\tau$ by fusing the flexibility of neural networks with the high-dimensional predictor space capabilities and extrapolation capability of the GPD model. Let the training dataset be $D = \{(x_i,y_i)\}_{i=1}^n$. Estimators for the intermediate quantile function $\tilde{Q}_x(\tau_0)$ with $\tau_0 < \tau$ and the GPD parameters $\sigma(x)$ and $\xi(x)$ are needed in order to estimate conditional extreme quantiles $\tilde{Q}_x(\tau)$ using \ref{eq11}. It is common practice to take two steps.  

First, they use traditional quantile regression techniques to estimate the intermediate quantile at level $\tau_0$. The conditional exceedances are then defined
\begin{equation*}
  z_i:=y_i-\tilde{Q}_{x_i}(\tau_0), \,\,\,\,\,\,\, i\in \emph{I}:=\{i=1,...,n: y_i>\tilde{Q}_{x_i}(\tau_0)\}
\end{equation*}
In order for the exceedances $z_i$ to represent approximate samples of a GPD, the intermediate probability $\tau_0$ should be selected so that it is large enough for the approximation in \ref{eq10} to be accurate but low enough to permit stable estimation of $Q_x(\tau_0)$ using standard empirical methods. However, since different values for $\tau_0$ result in distinct subsets of exceedances I, it is not a traditional tuning parameter. Thus, it would be meaningless to compare the loss function \ref{eq12} on these datasets. In the univariate scenario, the threshold is typically chosen based on sensitivity analysis and stability plots.
The GPD parameters $\sigma(x)$ and $\xi(x)$ are estimated in the second phase using the set of exceedances $z_i,\,\,\, i \in I$. Strong reliance between the estimates and numerical instabilities may result from directly modeling these factors in the extrapolation formula \ref{eq11}. Therefore, an orthogonal reparametrization with a diagonal Fisher information matrix is what we rely on. Regarding the usual asymptotic GPD likelihood characteristics, the reparametrization is well-defined for the GPD model for $\xi(x) > -0.5$
\begin{equation*}
  (\sigma(x),\xi(x))\mapsto(\nu(x),\xi(x)), \,\,\,\,\,\,\, \nu(x):=\sigma(x)(\xi(x)+1)
\end{equation*}
produces the necessary orthogonality (Chavez-Demoulin and Davison \cite{cha}; Cox and Reid \cite{co}). This reparametrization greatly enhances stability and convergence in all settings examined in the studies.
For the orthogonalized GPD parameters $\nu(x;W)$ and $\xi(x;W)$, where $W$ represents the collection of all model parameters, Pasche et al. \cite{pa} suggest a flexible neural network model. The parametric family is the GPD model with parameters $\theta = (\nu,\sigma)$ depending on the covariate $X = x$, and this may be viewed as conditional density estimation with output dimension $q = 2$. The GPD deviation loss across the training exceedances is often minimized by an estimate of the model parameters $\hat{W}$ is thus
\begin{equation*}
  \hat{W}\in arg_W min\sum_{i\in I} l_{OGPD}\{z_i;\hat{\nu}(x_i;W);\hat{\xi}(x_i;W)\}, 
\end{equation*}
where the orthogonal reparametrization's deviation or negative log-likelihood of the GPD is
\begin{equation}\label{eq12}
  l_{OGPD}(z;\nu,\xi)=\left(1+\frac{1}{\xi}\right)\log\left\{1+\xi\frac{(\xi+1)z}{\nu}\right\}+\log(\nu)-\log(\xi+1).
\end{equation}
Pasche et al. \cite{pa} discuss the model's specifics for both independent observations and sequentially dependent time series data. In this paper, we focus on the sequential dependence EQRN algorithm.\\

In the following section, we will conduct a simulation study to learn more about the prospects and limitations of each approach.

\subsection{Simulation results}
In this section, we compare the Carvalho et al. \cite{ca} method and the Pasche et al. \cite{pa} method by using the Root Mean Square Error (RMSE), Mean Absolute Error (MAE) metrics. Let the training data $\{(x_i,y_i)\}_{i=1}^N $ which are observed sequentially from a time series $\{(X_i,Y_i)\}_{i=1}^N$ with $N =7000$. In fact, based on all available data $\tilde{X}_u   =\{(X_t,Y_t)\}_{t<u}$, we would like to forecast the response $Y_u$ at a future time point $u$ as well as any potential high quantiles defined as 
\begin{equation}\label{eq13}
  Q_{\tilde{x}_u}(\tau):=F^{-1}_{Y_u|\tilde{X}_u=\tilde{x}_u}(\tau)
\end{equation}
where $\tilde{x}_u:=\{(x_t,y_t)\}_{t<u}$ are observations that may or may not be included in the training data. In the first, we give an overview of this algorithm. To precisely estimate extreme quantiles, the Extreme Quantile Regression Network (EQRN) technique integrates neural networks with extreme value theory. First, it models an intermediate quantile that acts as a threshold using conventional quantile regression. Second, it calculates the shape $\xi$ and scale $\sigma$, two parameters of the generalized Pareto distribution (GPD), as a function of the covariates. A neural network that understands the intricate relationships between predictors is used to generate these parameters. Lastly, the method extrapolates higher quantiles using GPD, which enables the prediction of uncommon events while accounting for spatial, temporal, or sequential information included in the data. Applications like flood or other extreme event predictions benefit greatly from this method; for more details, see Pasche et al. \cite{pa}.\\ For the Carvalho et al. \cite{ca} model, we use a typical componentwise adaptive Markov Chain Monte Carlo
 (MCMC)\cite{ha} with a Dirichlet prior, Dirichlet($10^{-4}\mathbf{1}_k$) defined on a generalized logit transformation
 of weights $\pi_\alpha$ in order to learn about regression lines from data. Every MCMC chain has a burn-in
 duration of 4 000 and a length of 10 000.

To accomplish this, we simulate the data in the following manner:
\begin{itemize}
  \item Scenario 1 : Let define the time series
\begin{equation*}
  X_t=0.4. X_{t-1}+|\zeta_t^X|,\,\,\,\,\,\zeta_t^X\sim \mathcal{N}(0,1)
\end{equation*}
and \begin{equation*}
      Y_t=\sigma_t|\zeta_t^Y|,\,\,\,\,\,\zeta_t^Y\sim \mathcal{N}(0,1)
    \end{equation*}
with 
\begin{equation*}
  \sigma_t^2=1+0.1\{2Y_{t-1}^2+Y_{t-2}^2+Y_{t-3}^2+Y_{t-4}^2+Y_{t-5}^2\}+0.1\{3X_{t-1}^2+2X_{t-2}^2+X_{t-3}^2+X_{t-4}^2+X_{t-5}^2\}
\end{equation*}
  \item Scenario 2 :  strongly dependent extremes: Husler-Reiss model with $\lambda = 0.1$;
  \item  Scenario 3 : weakly dependent extremes: Logistic model with $\alpha = 0.9$;
  \item Scenario 4 : asymmetric intermediate dependence: Coles–Tawn model with $\alpha = 0.5, \beta= 100$.
\end{itemize}
\begin{figure}[h!]
  \centering
  \begin{subfigure}{0.45\textwidth}
    \centering
    \includegraphics[width=\textwidth]{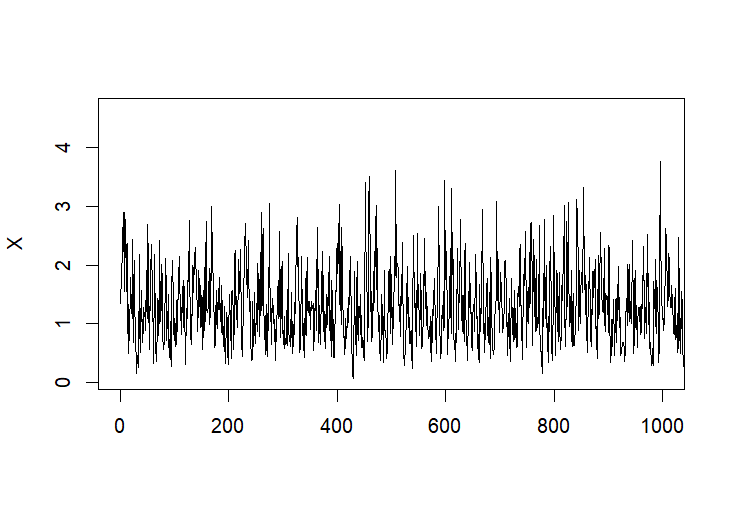}
    \caption{Series X}
  \end{subfigure}
  \begin{subfigure}{0.45\textwidth}
    \centering
    \includegraphics[width=\textwidth]{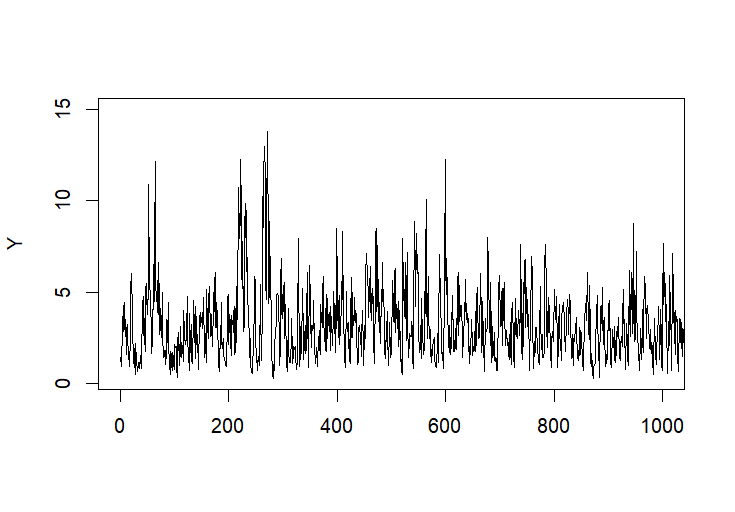}
    \caption{Series Y}
  \end{subfigure}
  \caption{1 000 observations of the Scenario 1 sequential data simulation}
  \label{fig:series}
\end{figure}

\begin{table}[h]
    \centering
    \begin{subtable}{0.45\textwidth}
        \centering
        \begin{tabular}{lcc}
            \hline
          & RMSE & MAE \\ \hline
            Carvalho et al. \cite{ca} & 509.76 & 127.72 \\ \hline
            Pasche et al. \cite{pa} & 1.65 & 1.23 \\ \hline
        \end{tabular}
        \caption{Scenario 1: Time series}
    \end{subtable}
    \begin{subtable}{0.45\textwidth}
        \centering
        \begin{tabular}{lcc}
            \hline
           & RMSE & MAE \\ \hline
            Carvalho et al. \cite{ca} & 755.18 & 243.7 \\ \hline
            Pasche et al. \cite{pa} & 0.04 & 0.03 \\ \hline
        \end{tabular}
        \caption{Scenario 2:  Husler-Reiss model }
    \end{subtable}
    
    \vspace{0.5cm} % Espacement entre les lignes  
    
    \begin{subtable}{0.45\textwidth}
        \centering
        \begin{tabular}{lcc}
            \hline
            & RMSE & MAE \\ \hline
            Carvalho et al. \cite{ca} & 340.6 & 85.09 \\ \hline
            Pasche et al. \cite{pa} & 0.02 & 0.01 \\ \hline
        \end{tabular}
        \caption{Scenario 3: Logistic model }
    \end{subtable}
    \begin{subtable}{0.45\textwidth}
        \centering
        \begin{tabular}{lcc}
            \hline
    & RMSE & MAE \\ \hline
            Carvalho et al. \cite{ca} & 755.18 & 242.33 \\ \hline
            Pasche et al. \cite{pa} & 0.03 & 0.02 \\ \hline
        \end{tabular}
        \caption{Scenario 4:  Coles–Tawn model }
    \end{subtable}   
    
    \caption{Performance comparison between models in Scenarios (1-4)}
    \label{tab:four_tables}
\end{table}
\begin{figure}[h!]
  \centering
  \begin{subfigure}{0.45\textwidth}
    \centering
    \includegraphics[width=\textwidth]{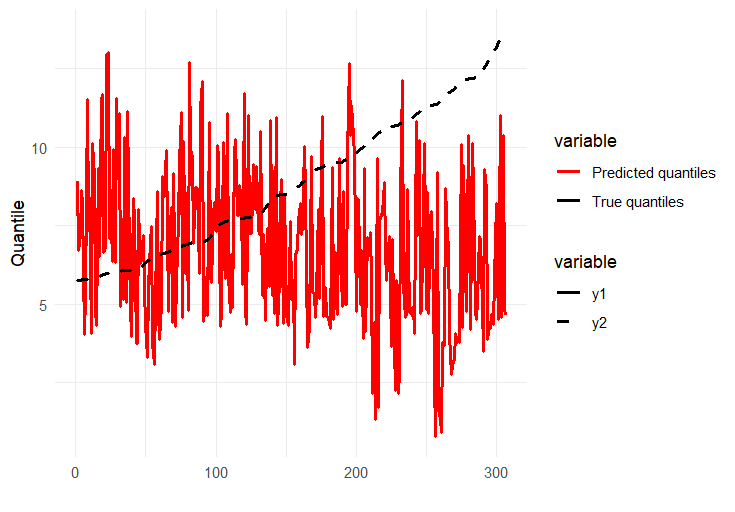}
    \caption{Series }
  \end{subfigure}
  \begin{subfigure}{0.45\textwidth}
    \centering
    \includegraphics[width=\textwidth]{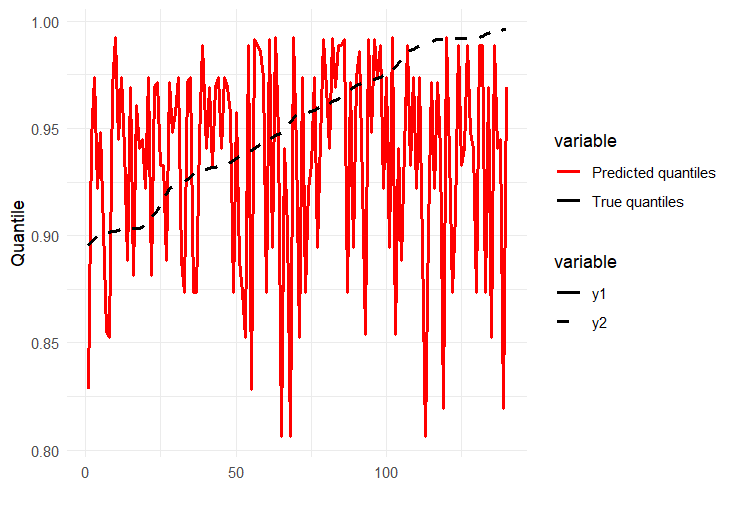}
    \caption{Husler-Reiss Model}
  \end{subfigure}
  \begin{subfigure}{0.45\textwidth}
    \centering
    \includegraphics[width=\textwidth]{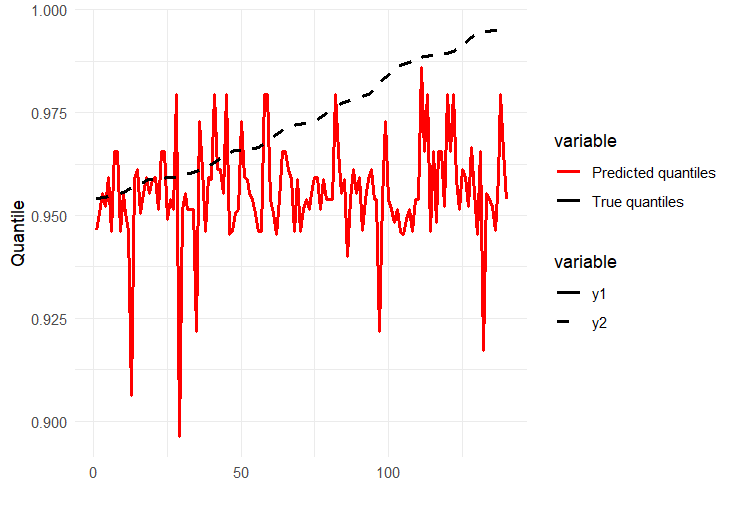}
    \caption{Logistic Model}
  \end{subfigure}
  \begin{subfigure}{0.45\textwidth}
    \centering
    \includegraphics[width=\textwidth]{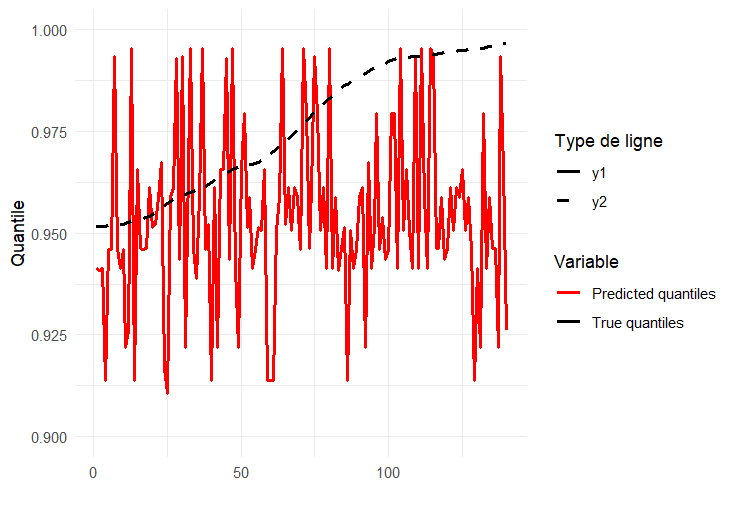}
    \caption{Coles-Tawn Model}
  \end{subfigure}
  \caption{True and 90\% predicted quantiles Pasche et al. \cite{pa}}
  \label{fig:quantiles}
\end{figure}

A portion of the simulated data is displayed in figure \ref{fig:series} . All approaches employ the same covariate vectors, $\tilde{x}_t=\{(x_j,y_j)\}_{j=t-s}^{t-1}$ with $s = 10$, to ensure a fair comparison. Since the conditional distribution $Y_t|\tilde{X}_t=\tilde{x}_t$ is a folded normal distribution, this model admits a GPD approximation as in equation \ref{eq10}. As a result, a GPD with shape parameter $\xi(\tilde{x}_t) = 0$ can approximate its tail.  \\

To accomplish this objective, we suggest a recurrent neural network in this section. In our study, the chosen model for intermediate quantile regression is a Generalized Random Forest (GRF) with 1000 trees, trained separately on the training and validation datasets with $\tau=90\%$. The algorithm EQRN for sequential data is employed to model the GPD parameters $\nu(\tilde{x}_t)$ and $\xi(\tilde{x}_t)$.We estimate high quantiles from a trained EQRN model for simulation using the EQRN Algorithm. Using intermediate quantiles to improve predictions, it applies the model to training data and makes quantile predictions at a 90\% probability level. A systematic method of sequential forecasting is ensured by setting the sequence length to 10. At a 90\% probability level, we estimate intermediate quantiles using Generalized Random Forest (GRF) quantile regression 1000 trees are used to train the model on both training and validation datasets. In order to improve the accuracy of the model, the projected quantiles are calculated out-of-bag, which means they are based on observations that were not utilized in the tree-building process. By capturing the behaviour of high quantiles in complicated datasets, these intermediate quantiles offer important statistical insights, especially for risk management and anticipating extreme events.\\

The table \ref{tab:four_tables} compare the performance of two models using the RMSE (Root Mean Squared Error) and MAE (Mean Absolute Error) metrics, which measure the mean squared error and the mean absolute error between predictions and actual values, respectively:\\
As shown in Scenario 1 table, with substantially lower errors (RMSE = 1.65, \\MAE = 1.23 vs. RMSE = 509.76, MAE = 127.72), the Pasche et al. \cite{pa} model outperforms the Carvalho et al. \cite{ca} model. This suggests better predictive accuracy and a significant decrease in the differences between expected and observed values.\\
According to the Husler-Reiss model table , the Pasche et al. \cite{pa} model demonstrates significantly better performance than the Carvalho et al. \cite{ca} model, with much lower errors (RMSE = 0.04, MAE = 0.03 vs. RMSE = 755.18, MAE = 243.7), indicating superior predictive accuracy and a substantial reduction in discrepancies between predicted and observed values. \\
With significantly lower errors (RMSE = 0.02, MAE = 0.01 vs. RMSE = 340.6, \\MAE = 85.09), the Pasche et al. \cite{pa} model outperforms the Carvalho et al. \cite{ca} model via the equation \ref{eqlog}, indicating superior predictive accuracy and a significant decrease in the discrepancies between predicted and observed values, as shown in the Logistic model table.\\
The Pasche et al. \cite{pa} model outperforms the Carvalho et al. \cite{ca} model by a significant margin, as shown in the Coles-Tawn Model table. The errors are significantly lower (RMSE = 0.03, MAE = 0.02 vs. RMSE = 755.18, MAE = 242.33), indicating superior predictive accuracy and a significant decrease in the differences between predicted and observed values.
RMSE (Root Mean Squared Error) and MAE (Mean Absolute Error), which quantify the squared and absolute discrepancies between predictions and actual values, are used in the tables to compare the performance of two models. With noticeably lower RMSE and MAE values, the Pasche et al. \cite{pa} model continuously performs better than the Carvalho et al. \cite{ca} model throughout all four tables. With a significant decrease in the differences between expected and observed values, this suggests that the Pasche et al. \cite{pa} model produces predictions that are more accurate and trustworthy.\\
One encouraging finding from this figure \ref{fig:quantiles} is that all models can approach the overall behaviour of the extremes by capturing the genuine quantiles' general rising tendency, especially in the upper range. The projected quantiles, in spite of some deviations, are still rather near to the genuine quantiles in a number of places, indicating that these models can still offer useful information for estimating high quantiles, which is essential for risk management and severe event prediction applications.
\subsection{Real data}
Now, we use our research to calculate the fire danger in Morocco's Fez-Meknes region. For this, we use the temperature measured at five sites (Fez, Boulemane, Sefrou, Ifrane, and Meknes-El Hajeb) as an explanatory covariate $X_p$ with $p=5$, and the number of fires as a response variable $Y_p$.\\ The data collection includes 60 observations of the highest yearly temperature from 2011 to 2022 from these five sites and 427 daily observations of fires. The covariates $x_t\in\mathbb{R}^p$, $p = 5$ show the highest temperature from five fire-threatened locations, while the response $y_t$ shows the number of fires per site.
The Fez-Meknes region (on the map above) is located in the north of Morocco, bordering Tanger-Tetouan-Al Hoceïma to the north, Orientale to the east, Drâa-Tafilalet to the south, Beni Mellal-Khenifra to the south-west and Rabat-Sale-Kenitra to the west. The Fez-Meknes region lies at coordinates $34^\circ\, 02'\, 00''$ north, $5^\circ \,00' \,00''$ west.\\
\begin{figure}[h]
  \centering
  \includegraphics[width=0.6\textwidth]{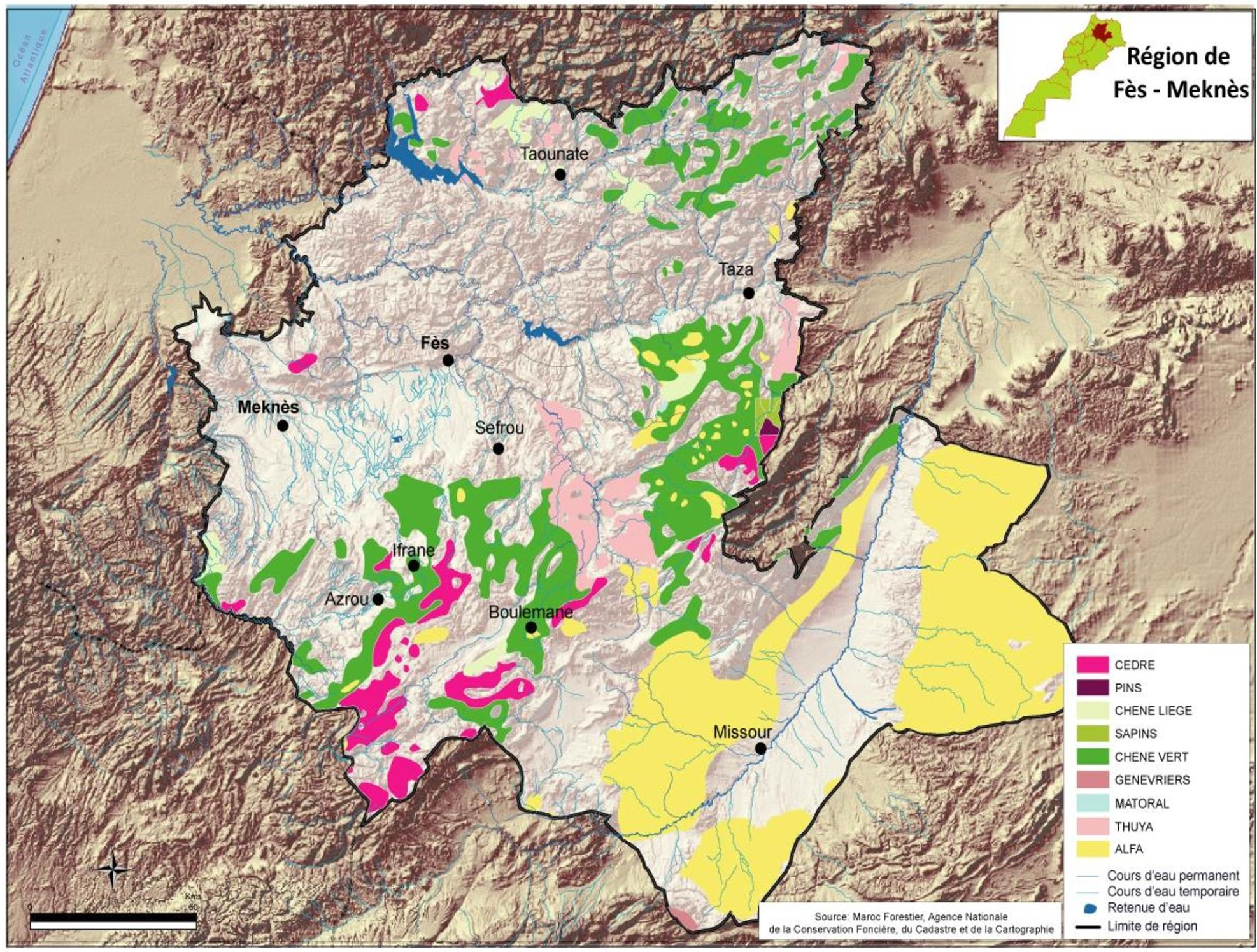}
  \caption{Forest cover map of Morocco's Fez-Meknes region}\label{map}
\end{figure}
The National Water and Forestry Agency's Fez-Meknes Regional Office provided the information on the number of fires at each station per year during the years 2011-2023. We gathered each station's highest annual temperature. Therefore, our study's goal is to forecast risk based on temperature using our model. \\ 
The models are trained using the initial T = 203 observations from 2011–2016, of which the first three quarters are utilized for parameter estimation and the remaining quarter is used as a validation set to identify hyperparameters (sets $\mathcal{T}$ and $\mathcal{V}$ in Algorithm 2, respectively). The test set, which includes 224 observations from 2017–2022, is only used to assess the model's performance across a separate time period; it is not utilized for parameter selection or fitting. We are gathering 60 temperature measurements from these five sites between 2011 and 2022, displaying annual maximums.\\
\begin{figure}[h]
  \centering
  \includegraphics[width=0.8\textwidth]{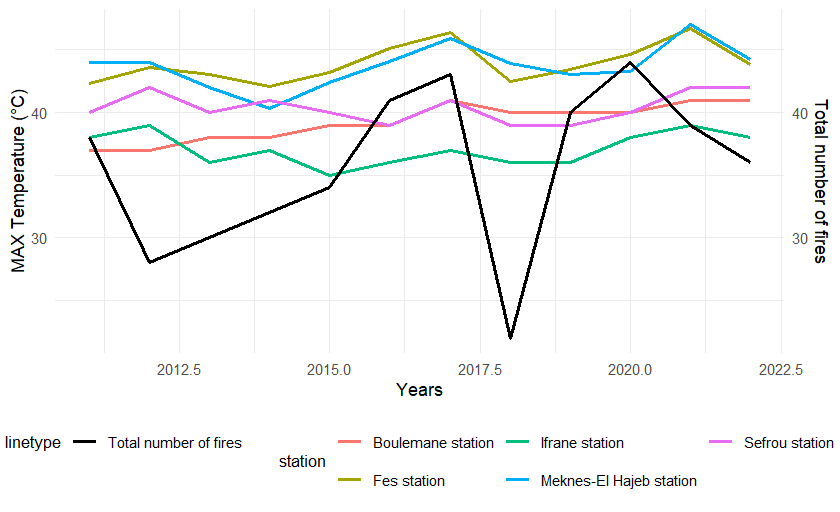}
  \caption{Maximum Temperatures and Total Number of Fires Across Stations Over Years 2011-2023 of Morocco’s Fez-Meknes region.}\label{data}
\end{figure}
First, let's look at the raw data: the association between the total number of fires from 2011 to 2022 and the highest temperatures recorded at five meteorological stations—Boulemane, Ifrane, Sefrou, Fez, and Meknes-El Hajeb—is depicted in the figure \ref{data}. Fes and Meknes-El Hajeb continuously record the highest temperatures, whereas the other stations show varying but generally high temperatures. The black line, which represents the overall number of fires, has discernible peaks in years with higher temperatures, especially in 2016, 2017, and 2020, pointing to a potential link between rising temperatures and an increase in fire events. On the other hand, 2018 had a notable decline in flames that correlated with a modest drop in temperature. This pattern suggests that rising temperatures could be a factor in the increased frequency of wildfires in the area.\\
A QRN with one LSTM layer of dimension 64, the standard fully connected layer, and an $L_2$ weight penalty with parameter $\lambda=10^{-4}$ is the final model selected to regress the intermediate quantiles. One LSTM layer of dimension 64, a fully connected layer, and an $L_2$ weight penalty with parameter $\lambda =10^{-4}$ are features of the selected EQRN model.

 According to Pasche et al. \cite{pa}, the $T$-year return level $Q^T$, or the magnitude of an event that is typically exceeded once every $T$ years, is frequently used in hydrology and climate science to evaluate risk. The quantile $Q^T = Q(1-1/(n_{Y}T))$ is the $T$-year return level if $Y$ is a quantity with $n_Y$ independent records annually (for example, $n_Y = 365$ for daily data and $n_Y = 1$ for yearly maximum).\\
 
We are focusing on the $Q^{100}$ quantile in this instance similarly to Pasche et al. \cite{pa}. According to the study, the Carvalho et al. \cite{ca} approach is more effective. We use it to model fire risk for this reason. For this, we use the maximum yearly temperatures gathered from five stations—Ifrane Station, Boulemane Station, Fez Station, Sefrou Station, and Meknes-El Hajeb Station—as covariates $\mathcal{X}_p$ with $p=5$ and the total number of fires as a response variable $\mathcal{Y}_i$.  For the analysis, we use observations for which $\hat{X}_i+\hat{Y}_i > u$ , where $u$ is the $98\%$ quantile; here, 
the raw data are transformed to unit Fréchet margins via the transformation $(\hat{X}_i, \hat{Y}_i)=(-1/\log\{\hat{F}_X(\mathcal{X}_i)\},-1/\log\{\hat{F}_Y(\mathcal{Y}_i)\})$, where $\hat{F}_X$ and $\hat{F}_Y$ respectively denote the empirical distribution functions (normalized by $n + 1$ rather than by $n$ to avoid division by zero).  
Solve the equation $G_{Y|X}(y|x)=q$ for $q=0.99$ that is the $Q^{100}$ to determine the conditional quantile, see the equations \ref{eq5} and \ref{eq9} . 
According to the study, the number of fires tends to rise by roughly $2.06$ in the higher quantile as the highest temperature at the sites rises. This suggests a strong correlation between extreme temperatures and fire occurrences, highlighting the increased risk in periods of intense heat. Such findings are crucial for developing predictive models and informing wildfire prevention strategies, especially in regions prone to climate-related disasters.\\

 \begin{table}
   \centering
   \begin{tabular}{lcc}
            \hline
    & RMSE & MAE \\ \hline
            Carvalho et al. \cite{ca} & 35.55 & 35.02 \\ \hline
            Pasche et al. \cite{pa} & 2.27 & 2.1 \\ \hline
        \end{tabular}
       
   \caption{Performance comparison of models in terms of the number of fires at the highest temperature each year  }\label{tab2}
 \end{table}
 Using a trained EQRN model, we use the EQRN Algorithm to simulate the estimation of extreme quantiles at a 95\% probability level. Using intermediate quantiles to improve predictions, it generates quantiles for the training dataset. We estimate extreme quantiles at a 95\% probability level using Generalized Random Forest (GRF) quantile regression. 1000 trees are used to train the model on both training and validation datasets, guaranteeing reliable predictions while preserving reproducibility. In order to improve the accuracy of the model, the projected quantiles are calculated out-of-bag, which means they are based on observations that were not utilized in the tree-building process. \\
 
  According to the table \ref{tab2}, the Pasche et al. \cite{pa} model exhibits much lower error values (RMSE = 2.27, MAE = 2.1) than the Carvalho et al. \cite{ca} model (RMSE = 35.55, MAE = 35.02), suggesting that the Pasche et al. \cite{pa} model has a lower departure from real values and better predicted accuracy. Carvalho et al.'s \cite{ca}model's high error values imply that additional optimization or an alternative strategy would be needed to increase reliability.\\
  \begin{figure}[h]
  \centering
  \includegraphics[width=0.8\textwidth]{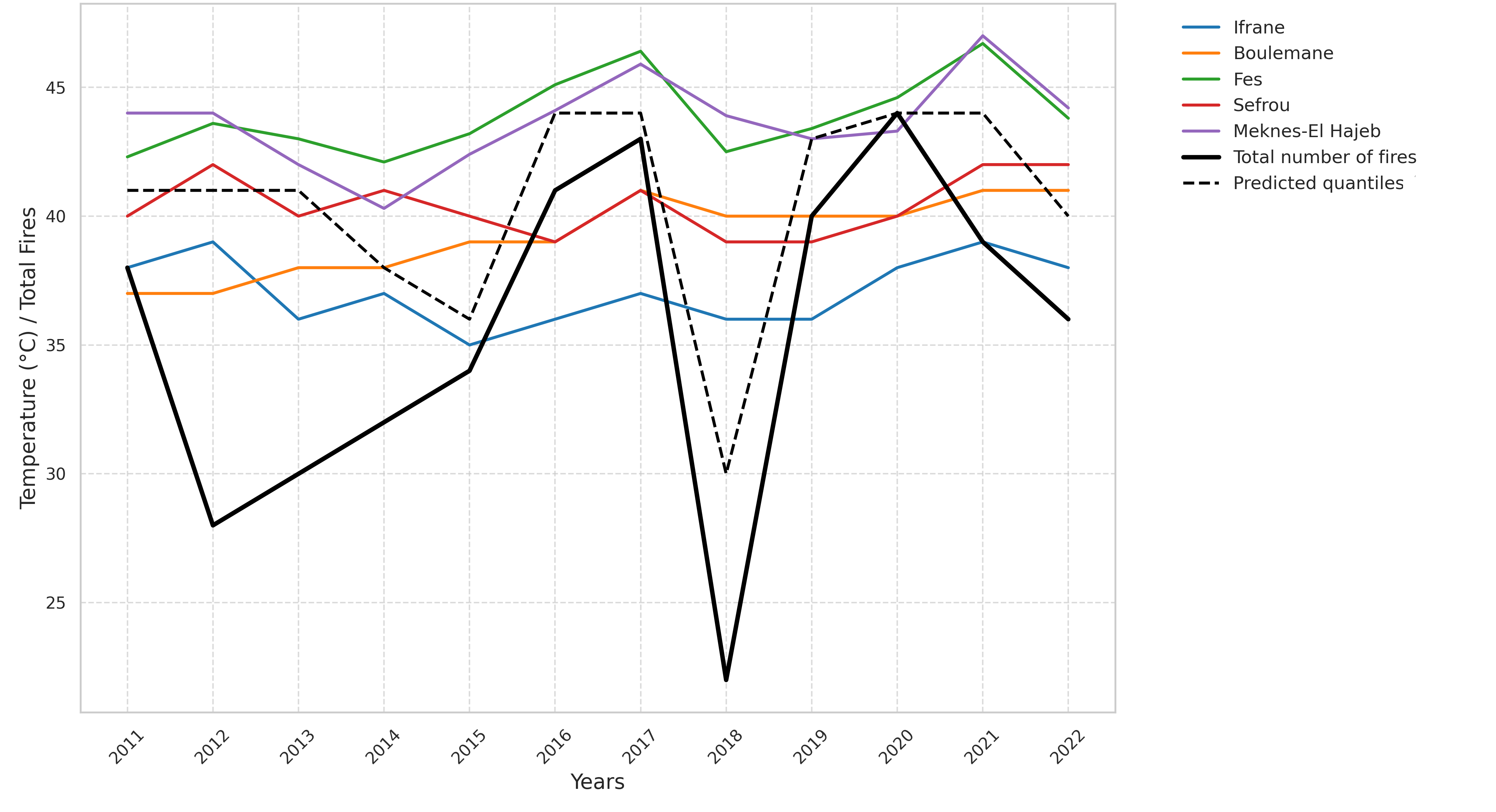}
  \caption{Maximum Temperatures and Total Number of Fires Across Stations with Predicted Quantiles (95\%) Over Years 2011-2023 of Morocco’s Fez-Meknes region.}\label{dataquantile}
\end{figure}
The figure \ref{dataquantile} illustrates the annual evolution of maximum temperatures recorded at five Moroccan stations (Ifrane, Boulemane, Fes, Sefrou, Meknes-El Hajeb) alongside the total number of fires and the predicted 95\% quantiles of fire occurrences from 2011 to 2022. Overall, the maximum temperatures show moderate fluctuations over the years, with Fes and Meknes-El Hajeb consistently registering the highest peaks, particularly in 2016, 2017, 2020, and 2021. The total number of fires (black line) tends to increase during years with higher temperature peaks, such as 2016, 2017, 2020, and 2021, highlighting a potential link between extreme temperatures and fire risk. The predicted quantiles (dashed line) follow a similar pattern to the observed fires, providing a reasonably good estimation of fire occurrence based on temperature trends, though with some discrepancies in years like 2018 and 2022, where actual fires were lower than predicted. This suggests that while temperature is a significant driver of fire risk, other environmental or human factors likely influence the variability in fire occurrences.
\section{Discussion}\label{sec3}
 
In the context of predicting fire danger in the Fez-Meknes region of Morocco, this paper offers a comparative examination of various quantile regression strategies for modelling extreme values. Utilizing extreme value theory, machine learning models, and Monte Carlo simulations, the study assesses the advantages and disadvantages of different methods for estimating high quantiles. The results demonstrate the superiority of the EQRN model, which combines neural networks and the generalized Pareto distribution (GPD) to provide robust extrapolation beyond historical data and improved predictive accuracy.\\

According to the simulation results, EQRN consistently performs better than conventional regression models, such the Carvalho et al. model, across a range of dependence patterns, such as asymmetric intermediate reliance (Coles-Tawn model) and very dependent extremes (Hüsler-Reiss model). The measures of Mean Absolute Error (MAE) and Root Mean Square Error (RMSE) verify that EQRN offers noticeably fewer mistakes, signifying better predictive ability. The relevance of quantile-based forecasting for risk management is further supported by the real-world application of these models to fire risk prediction, which shows a positive association between increasing maximum temperatures and fire frequency.\\

In this regard, we might consider novel neural network-based models that may improve the capacity to grasp intricate relationships in extreme occurrences. A useful approach for enhancing quantile estimation is neural networks, especially deep learning architectures, which have demonstrated encouraging outcomes in risk modelling and time series forecasting. The study promotes a hybrid modelling method to improve the interpretability and accuracy of severe event predictions by fusing machine learning flexibility with statistical rigour. The study shows how intermediate quantiles can improve predictions and guarantee reliable sequential forecasting by combining EQRN with Generalized Random Forest (GRF) quantile regression. According to the findings, quantile regression neural networks are a potent tool for making decisions in unpredictable situations and can be successfully used for financial risk modelling, climate risk assessment, and natural catastrophe prediction. Future research could explore hybrid architectures, combining deep learning techniques with statistical extreme value models, to improve forecasting accuracy and interpretability in high-risk scenarios.
\section*{Data availability}
The datasets analyzed in this study can be accessed from TuTiempo.net and Historique-Meteo.net free web
sites’ : these data are downloaded from \url{https://fr.tutiempo.net/} and \url{https://www.historique-meteo.net/afrique/maroc/}, which have authorized us to use them.
\section*{Acknowledgments}
We warmly thank all those who contributed to this research, especially Mr. Mohamed Chofqi and the institutions providing the meteorological data, as well as the research teams for their valuable analyses and discussions. Their support and expertise greatly enriched this study.
\section*{ Conflicts of Interest}
 The authors declare that they have no conflicts of Interest.

\end{document}